\def\year{2020}\relax
\documentclass[letterpaper]{article} 
\usepackage{aaai20}  
\usepackage{times}  
\usepackage{helvet} 
\usepackage{courier}  
\usepackage[hyphens]{url}  
\usepackage{graphicx} 
\usepackage{graphicx}  
\usepackage{subcaption}
\usepackage[table,x11names]{xcolor}
\usepackage{pifont}
\usepackage{tabularx}
\usepackage{graphicx}
\usepackage {tikz}
\usetikzlibrary {positioning}
\usepackage{amsmath}
\usepackage{enumitem}

\newcommand{\citet}[1]{\citeauthor{#1} \shortcite{#1}}
\newcommand{\citep}{\cite}

\newcolumntype{L}{>{\centering\arraybackslash}m{3cm}}

\title{Aggressive, Repetitive, Intentional, Visible, and Imbalanced: \\ Refining Representations for Cyberbullying Classification}
 \author{Caleb Ziems\\
{Emory University}\\
cjziems@gmail.com
\And
Ymir Vigfusson\\
{Emory University}\\
ymir@mathcs.emory.edu
\And
Fred Morstatter\\
{USC Information Sciences Institute}\\
fredmors@isi.edu}

\begin{document}

\maketitle

\begin{abstract}
Cyberbullying is a pervasive problem in online communities. To identify cyberbullying cases in large-scale social networks, content moderators depend on machine learning classifiers for automatic cyberbullying detection. However, existing models remain unfit for real-world applications, largely due to a shortage of publicly available training data and a lack of standard criteria for assigning ground truth labels. In this study, we address the need for reliable data using an original annotation framework. Inspired by social sciences research into bullying behavior, we characterize the nuanced problem of cyberbullying using five explicit factors to represent its social and linguistic aspects. We model this behavior using social network and language-based features, which improve classifier performance. These results demonstrate the importance of representing and modeling cyberbullying as a social phenomenon. 
\end{abstract}

\section{Introduction}
Cyberbullying poses a serious threat to the safety of online communities.  The Centers for Disease Control and Prevention (CDC) identify cyberbullying as a ``growing public health problem in need of additional research and prevention efforts'' \cite{david2009electronic}. Cyberbullying has been linked to negative mental health outcomes, including depression, anxiety, and other forms of self-harm, suicidal ideation, suicide attempts, and difficulties with social and emotional processing \cite{miller2016cyberbullying,price2010cyberbullying,sampasa2014associations}.
Where traditional bullying was once limited to a specific time and place, cyberbullying can occur at any hour and from any location on earth \cite{chatzakou2017mean}. Once the first message has been sent, the attack can escalate rapidly as harmful content is spread across shared media, compounding these negative effects \cite{waasdorp2015overlap,huang2010analysis}.

Internet users depend on content moderators to flag abusive text and ban cyberbullies from participating in online communities. However, due to the overwhelming volume of social media data produced every day, manual human moderation is often unfeasible. For this reason, social media platforms are beginning to rely instead on machine learning classifiers for automatic cyberbullying detection \cite{van2018automatic}. 

The research community has developed increasingly competitive classifiers to detect harmful or aggressive content in text. Despite significant progress in recent years, however, existing models remain unfit for real-world applications. This is due, in part, to shortcomings in the training and testing data \cite{hosseinmardi2016prediction,salawu2017approaches,rosa2019automatic}. Most annotation schemes have ignored the importance of social context, and researchers have neglected to provide annotators with objective criteria for distinguishing cyberbullying from other crude messages. 

\begin{table*}[t]
\centering
\caption{Datasets built from different related definitions of cyberbullying. For each dataset, we report the size, positive class balance, inter-annotator agreement, and whether the study incorporated social context in the annotation process.}
\resizebox{\textwidth}{!}{%
\begin{tabular}{l|ccccc|lrrr|c} 
\rowcolor{lightgray}\textbf{Work}&\textsc{aggr}&\textsc{rep}&\textsc{harm}&\textsc{peer}&\textsc{power} &
\textbf{Data Source}&\textbf{Size}&\textbf{Balance}&\textbf{Agreement}&\textbf{Context}\\ \hline
\citet{al2016cybercrime} & \ding{51} &  & \ding{51} &  & 
& Twitter & 10,007 & 6.0\% & -- & \ding{55}\\ \hline
\citet{chatzakou2017mean} & \ding{51} & \ding{51}  & \ding{51} &  & \ding{51} 
& Twitter & 9,484 & -- & 0.54 & \ding{51}\\ \hline
\citet{hosseinmardi2015analyzing} & \ding{51} & \ding{51} & & & \ding{51} 
& Instagram & 1,954 & 29.0\% & 0.50 & \ding{51} \\ \hline
\citet{huang2014cyber} &  & \ding{51} & \ding{51} & &  
& Twitter & 4,865 & 1.9\% & -- & \ding{55} \\ \hline
\citet{reynolds2011using} &  & \ding{51} & \ding{51} & &  
& Formspring & 3,915 & 14.2\% & -- & \ding{55}\\ \hline
\citet{rosa2019automatic} & \ding{51} & \ding{51}& \ding{51} & \ding{51} & 
& Formspring & 13,160 & 19.4\% & -- & \ding{55}\\ \hline
\citet{sugandhi2016automatic} & & \ding{51} & \ding{51} & &  
& Mixed & 3,279 & 12.0\% & -- & \ding{55}\\ \hline
\citet{van2018automatic} & \ding{51} & & \ding{51} &  &  
& AskFM & 113,698 & 4.7\% & 0.59 & \ding{51}
\end{tabular}%
}
\label{tab:def}
\label{tab:sets}
\end{table*}

To address the urgent need for reliable data, we provide an original annotation framework and an annotated Twitter dataset.\footnote{https://github.com/cjziems/cyberbullying-representations} The key advantages to our labeling approach are:
\begin{itemize}[leftmargin=.2in]
    \item \textbf{Contextually-informed ground truth.} We provide annotators with the social context surrounding each message, including the contents of the reply thread and the account information of each user involved.
    \item \textbf{Clear labeling criteria.} We ask annotators to provide labels for five clear cyberbullying criteria. These criteria can be combined and adapted for revised definitions of cyberbullying.
\end{itemize}
Using our new dataset, we experiment with existing NLP features and compare results with a newly-proposed set of features. We designed these features to encode the dynamic relationship between a potential bully and victim, using comparative measures from their relative linguistic and social network profiles. Additionally, our features have low computational complexity, so they can scale to internet-scale datasets, unlike expensive network centrality and clustering measurements.

Results from our experiments suggest that, although existing NLP models can reliably detect aggressive language in text, these lexically-trained classifiers will fall short of the more subtle goal of cyberbullying detection. With $n$-grams and dictionary-based features, classifiers prove unable to detect harmful intent, visibility among peers, power imbalance, or the repetitive nature of aggression with sufficiently high precision and recall. However, our proposed feature set improves $F_1$ scores on all four of these social measures. Real-world detection systems can benefit from our proposed approach, incorporating the social aspects of cyberbullying into existing models and training these models on socially-informed ground truth labels.

\section{Background}
Existing approaches to cyberbullying detection generally follow a common workflow. Data is collected from social networks or other online sources, and ground truth is established through manual human annotation. Machine learning algorithms are trained on the labeled data using the message text or hand-selected features. Then results are typically reported using precision, recall, and $F_1$ scores. Comparison across studies is difficult, however, because the definition of cyberbullying has not been standardized. Therefore, an important first step for the field is to establish an objective definition of cyberbullying.

\subsection{Defining Cyberbullying}
Some researchers view cyberbullying as an extension of more ``traditional'' bullying behaviors \cite{hinduja2008cyberbullying,olweus2012cyberbullying,raskauskas2007involvement}. In one widely-cited book, the psychologist Dan Olweus defines schoolyard bullying in terms of three criteria: \textbf{repetition}, \textbf{harmful intent}, and an \textbf{imbalance of power} \cite{olweus1994bullying}. He then identifies bullies by their intention to ``inflict injury or discomfort'' upon a weaker victim through repeated acts of aggression.

Social scientists have extensively studied this form of bullying as it occurs among adolescents in school \cite{kowalski2013psychological,li2006cyberbullying}. However, experts disagree whether cyberbullying should be studied as a form of traditional bullying or a fundamentally different phenomenon \cite{kowalski2013psychological,olweus2012cyberbullying}. Some argue that, although cyberbullying might involve repeated acts of aggression, this condition might not necessarily hold in all cases, since a single message can be otherwise forwarded and publicly viewed without repeated actions from the author \cite{slonje2013nature,waasdorp2015overlap}. Similarly, the role of power imbalance is uncertain in online scenarios. Power imbalances of physical strength or numbers may be less relevant, whereas bully anonymity and the permanence of online messages may be sufficient to render the victim defenseless \cite{slonje2008cyberbullying}. 

The machine learning community has not reached a unanimous definition of cyberbullying either. They have instead echoed the uncertainty of the social scientists. Moreover, some authors have neglected to publish any objective cyberbullying criteria or even a working definition for their annotators, and among those who do, the formulation varies. This disagreement has slowed progress in the field, since classifiers and datasets cannot be as easily compared. Upon review, however, we found that all available definitions contained a strict subset of the following criteria: aggression (\textsc{aggr}), repetition (\textsc{rep}), harmful intent (\textsc{harm}), visibility among peers (\textsc{peer}), and power imbalance (\textsc{power}). The datasets built from these definitions are outlined in Table~\ref{tab:def}. 

\subsection{Existing Sources of Cyberbullying Data}

According to \citet{van2018automatic}, data collection is the most restrictive ``bottleneck'' in cyberbullying research. Because there are very few publicly available datasets, some researchers have turned to crowdsourcing using Amazon Mechanical Turk or similar platforms.

In most studies to date, annotators labeled individual messages instead of message threads, ignoring social context altogether~\cite{al2016cybercrime,huang2014cyber,nahar2014semi,reynolds2011using,singh2016cyberbullying,sugandhi2016automatic}. Only three of the papers that we reviewed incorporated social context in the annotation process. \citet{chatzakou2017mean} considered batches of time-sorted tweets called \textit{sessions}, which were grouped by user accounts, but they did not include message threads or any other form of context. \citet{van2018automatic} presented ``original conversation[s] when possible,'' but they did not explain when this information was available. \citet{hosseinmardi2016prediction} was the only study to label full message reply threads as they appeared in the original online source. 

\subsection{Modeling Cyberbullying Behavior}
A large body of work has been published on cyberbullying detection and prediction, primarily through the use of natural language processing techniques. Most common approaches have relied on lexical features such as $n$-grams \cite{hosseinmardi2016prediction,van2018automatic,xu2012learning}, TF-IDF vectors \cite{dinakar2011modeling,nahar2013cyberbullying,sugandhi2016automatic}, word embeddings \cite{zhao2016automatic}, or phonetic representations of messages \cite{zhang2016cyberbullying}, as well as dictionary-based counts on curse words, hateful or derogatory terms, pronouns, emoticons, and punctuation \cite{al2016cybercrime,dadvar2013improving,reynolds2011using,singh2016cyberbullying}. Some studies have also used message sentiment \cite{singh2016cyberbullying,sugandhi2016automatic,van2018automatic} or the age, gender, personality, and psychological state of the message author according to text from their timelines \cite{al2016cybercrime,dadvar2013improving}. These methods have been reported with appreciable success as shown in Table~\ref{tab:sota}. 

Some researchers argue, however, that lexical features alone may not adequately represent the nuances of cyberbullying. \citet{hosseinmardi2015analyzing} found that among Instagram media sessions containing profane or vulgar content, only 30\% were acts of cyberbullying. They also found that while cyberbullying posts contained a moderate proportion of negative terms, the most negative posts were not considered cases of cyberbullying by the annotators. Instead, these negative posts referred to politics, sports, and other domestic matters between friends \cite{hosseinmardi2015analyzing}. 

The problem of cyberbullying cuts deeper than merely the exchange of aggressive language. The meaning and intent of an aggressive post is revealed through conversation and interaction between peers. Therefore, to properly distinguish cyberbullying from other uses of aggressive or profane language, future studies should incorporate key indicators from the social context of each message. Specifically, researchers can measure the author's status or social advantage, the author's harmful intent, the presence of repeated aggression in the thread, and the visibility of the thread among peers \cite{hosseinmardi2015analyzing,rosa2019automatic,salawu2017approaches}. 

Since cyberbullying is an inherently social phenomenon, some studies have naturally considered social network measures for classification tasks. Several features have been derived from the network representations of the message interactions. The degree and eigenvector centralities of nodes, the $k$-core scores, and clustering of communities, as well as the tie strength and betweenness centralities of mention edges have all been shown to improve text-based models \cite{huang2014cyber,singh2016cyberbullying}. Additionally, bullies and victims can be more accurately identified by their relative network positions. For example, the Jaccard coefficient between neighborhood sets in bully and victim networks has been
found to be statistically significant \cite{chelmis2017mining}. The ratio of all messages sent and received by each user was also significant.

These findings show promising directions for future work. Social network features may provide the information necessary to reliably classify cyberbullying. However, it may be prohibitively expensive to build out social networks for each user due to time constraints and the limitations of API calls \cite{yao2019cyberbullying}. For this reason, alternative measurements of online social relationships should be considered.

In the present study, we leverage prior work by incorporating linguistic signals into our classifiers. We extend prior work by developing a dataset that better reflects the definitions of cyberbullying presented by social scientists, and by proposing and evaluating a feature set that represents information pertaining to the social processes that underlie cyberbullying behavior.

\begin{table}
\centering
\caption{State of the Art in Cyberbullying Detection. Here, results are reported on either the Cyberbullying (CB) class exclusively or on the entire (total) dataset.}
\resizebox{\columnwidth}{!}{%
\begin{tabular}{Llrrrc} 
\rowcolor{lightgray}\textbf{Work}&\textbf{Model}&\textbf{Precision}&\textbf{Recall}&\textbf{F1}&\textbf{Class}\\ \hline
\citet{zhang2016cyberbullying} & CNN & 99.1\% & 97.0\% & 98.0\% & total\\ \hline
\citet{al2016cybercrime} & Random Forest & 94.1\% & 93.9\% & 93.6\% & total \\ \hline
\citet{nahar2014semi} & SVM & 87.0\% & 97.0\% & 92.0\% & CB \\ \hline \citet{sugandhi2016automatic} & SVM & 91.0\% & 91.0\% & 91.0\% & total\\ \hline
\citet{soni2018time} & Na\"ive Bayes & 80.2\% & 80.2\% & 80.2\% & total\\ \hline
\citet{zhao2016automatic} & SVM & 76.8\% & 79.4\% & 78.0\% & total\\ \hline
\citet{xu2012learning} & SVM & 76.0\% & 79.0\% & 77.0\% & total\\ \hline
\citet{hosseinmardi2016prediction} & Logistic Regression & 78.0\% & 72.0\% & 75.0\% & CB \\ \hline
\citet{yao2019cyberbullying} & CONcISE & 69.5\% & 79.4\% & 74.1\% & CB\\ \hline
\citet{van2018automatic} & SVM & 73.3\% & 57.2\% & 64.3\% & total\\ \hline
\citet{singh2016cyberbullying} & Proposed & 82.0\% & 53.0\% & 64.0\% & CB \\ \hline
\citet{rosa2019automatic} & SVM & 46.0\% & - & 45.0\% & CB \\ \hline
\citet{dadvar2013improving} & SVM & 31.0\% & 15.0\% & 20.0\% & CB \\ \hline
\citet{huang2014cyber} & Dagging & 76.3\% & - & - & CB 
\end{tabular}%
}
\label{tab:sota}
\end{table}

\section{Curating a Comprehensive\\ Cyberbullying Dataset}

Here, we provide an original annotation framework and a new dataset for cyberbullying research, built to unify existing methods of ground truth annotation. In this dataset, we decompose the complex issue of cyberbullying into five key criteria, which were drawn from the social science and machine learning communities. These criteria can be combined and adapted for revised definitions of cyberbullying. 

\subsection{Data Collection}
We collected a sample of 1.3 million unlabeled tweets from the Twitter Filter API. Since cyberbullying is a social phenomenon, we chose to filter for tweets containing at least one ``@'' mention. To restrict our investigation to original English content, we removed all non-English posts and retweets (RTs), narrowing the size of our sample to 280,301 tweets.

Since aggressive language is a key component of cyberbullying~\cite{hosseinmardi2015analyzing}, we ran the pre-trained classifier of \citet{davidson2017automated} over our dataset to identify hate speech and aggressive language and increase the prevalence of cyberbullying examples \footnote{Without this step, our positive class balance would be prohibitively small. See Appendix 1 for details.}. This gave us a filtered set of 9,803 aggressive tweets.

We scraped both the user and timeline data for each author in the aggressive set, as well as any users who were mentioned in one of the aggressive tweets. In total, we collected data from 21,329 accounts. For each account, we saved the full user object, including profile name, description, location, verified status, and creation date. We also saved a complete list of the user's friends and followers, and a 6-month timeline of all their posts and mentions from January $1^\text{st}$ through June $10^\text{th}$, 2019. For author accounts, we extended our crawl to include up to four years of timeline content. Lastly, we collected metadata for all tweets belonging to the corresponding message thread for each aggressive message.

\subsection{Annotation Task}
We presented each tweet in the dataset to three separate annotators as a Human Intelligence Task (HIT) on Amazon's Mechanical Turk (MTurk) platform. By the time of recruitment, 6,897 of the 9,803 aggressive tweets were accessible from the Twitter web page. The remainder of the tweets had been removed, or the Twitter account had been locked or suspended. 

We asked our annotators to consider the full message thread for each tweet as displayed on Twitter's web interface. We also gave them a list of up to 15 recent mentions by the author of the tweet, directed towards any of the other accounts mentioned in the original thread. Then we asked annotators to interpret each tweet in light of this social context, and had them provide us with labels for five key cyberbullying criteria. We defined these criteria in terms of the \emph{author} account (``who posted the given tweet?'') and the \emph{target} (``who was the tweet about?'' -- not necessarily the first mention). We also stated that ``if the target is not on Twitter or their handle cannot be identified'' the annotator should ``please write \textit{OTHER}.'' With this framework established, we gave the definitions for our five cyberbullying criteria as follows.
\begin{enumerate}[leftmargin=.2in]
	\item \textbf{Aggressive language:} (\textsc{aggr}) Regardless of the author's intent, the language of the tweet could be seen as aggressive. The user either addresses a group or individual, and the message contains at least one phrase that could be described as \textit{confrontational, derogatory, insulting, threatening, hostile, violent, hateful,} or \textit{sexually abusive}.
	\item \textbf{Repetition:} (\textsc{rep}) The target user has received at least two aggressive messages in total (either from the author or from another user in the visible thread).
	\item \textbf{Harmful intent:} (\textsc{harm}) The tweet was designed to tear down or disadvantage the target user by causing them distress or by harming their public image. The target does not respond agreeably as to a joke or an otherwise lighthearted comment.
	\item \textbf{Visibility among peers:} (\textsc{peer}) At least one other user besides the target has liked, retweeted, or responded to at least one of the author's messages.
	\item \textbf{Power imbalance:} (\textsc{power}) Power is derived from authority and perceived social advantage. Celebrities and public figures are more powerful than common users. Minorities and disadvantaged groups have less power. Bullies can also derive power from peer support.
\end{enumerate}
Each of these criteria was represented as a binary label, except for \textbf{power imbalance}, which was ternary. We asked ``Is there strong evidence that the \textbf{author} is more powerful than the target? Is the \textbf{target} more powerful? Or if there is not any good evidence, just mark \textbf{equal}.'' We recognized that an imbalance of power might arise in a number of different circumstances. Therefore, we did not restrict our definition to just one form of power, such as follower count or popularity.

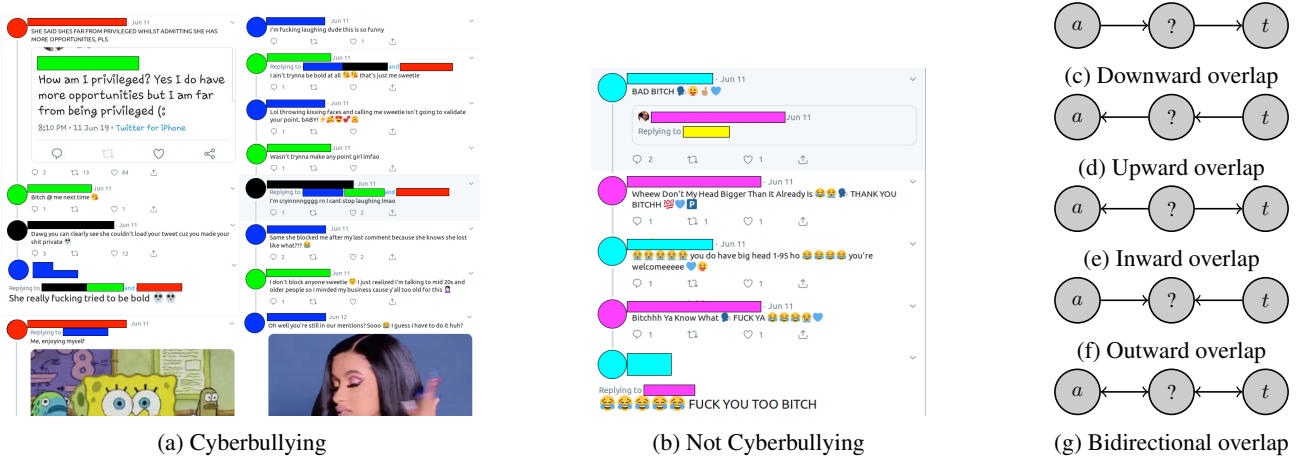
\begin{figure*}
    \centering
    \begin{subfigure}[b]{0.36\textwidth}
         \centering
         \includegraphics[width=\textwidth]{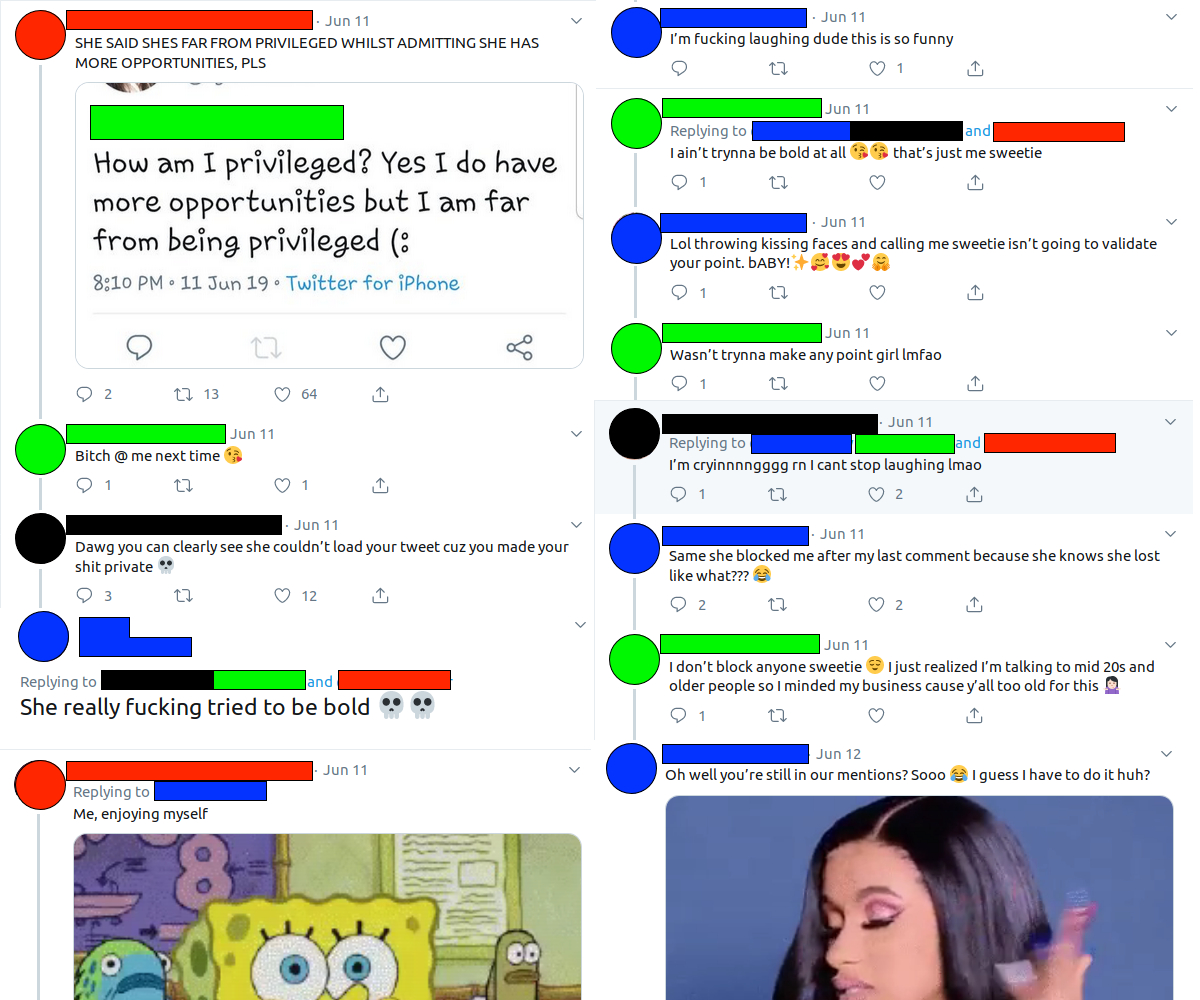}
         \caption{Cyberbullying}
         \label{fig:threadcb}
     \end{subfigure}
     \hspace{0.5in}
     \begin{subfigure}[b]{0.25\textwidth}
         \centering
         \includegraphics[width=\textwidth]{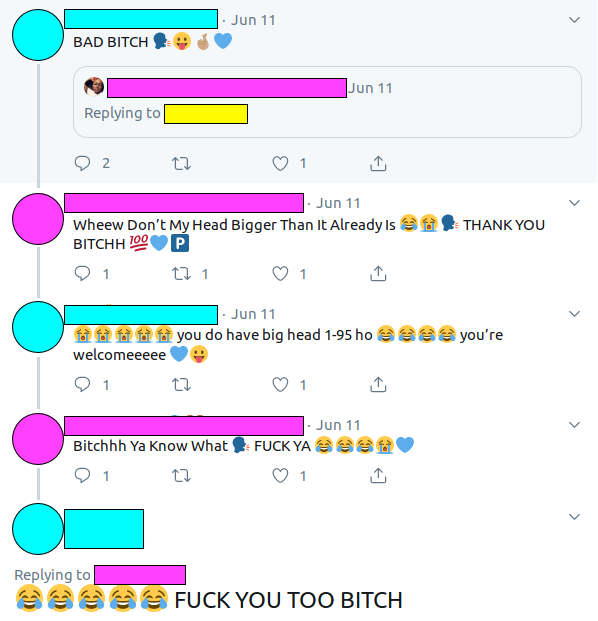}
         \caption{Not Cyberbullying}
         \label{fig:threadncb}
     \end{subfigure}
  \hfill
  \begin{subfigure}[b]{0.25\textwidth}
  {
    \centering
    \begin {tikzpicture}[every node/.style={scale=0.8}, node distance = 1cm and 1.25cm, on grid, semithick,
        state/.style ={circle, draw, fill=black!20, text=black, minimum width = 0.75cm}]
        \node[state] (A) {$a$};
        \node[state] (B) [right=of A] {?};
        \node[state] (C) [right=of B] {$t$};
        \path [->] (A) edge (B);
        \path [->] (B) edge (C);
    \end{tikzpicture}
    \caption{Downward overlap}
  
    \begin {tikzpicture}[every node/.style={scale=0.8},node distance = 1cm and 1.25cm, on grid, semithick,
        state/.style ={circle, draw, fill=black!20, text=black, minimum width = 0.75cm}]
        \node[state] (A) {$a$};
        \node[state] (B) [right=of A] {?};
        \node[state] (C) [right=of B] {$t$};
        \path [->] (B) edge (A);
        \path [->] (C) edge (B);
    \end{tikzpicture}
    \caption{Upward overlap}

    \begin {tikzpicture}[every node/.style={scale=0.8},node distance = 1cm and 1.25cm, on grid, semithick,
        state/.style ={circle, draw, fill=black!20, text=black, minimum width = 0.75cm}]
        \node[state] (A) {$a$};
        \node[state] (B) [right=of A] {?};
        \node[state] (C) [right=of B] {$t$};
        \path [->] (B) edge (A);
        \path [->] (B) edge (C);
    \end{tikzpicture}
    \caption{Inward overlap}

    \begin {tikzpicture}[every node/.style={scale=0.8},node distance = 1cm and 1.25cm, on grid, semithick,
        state/.style ={circle, draw, fill=black!20, text=black, minimum width = 0.75cm}]
        \node[state] (A) {$a$};
        \node[state] (B) [right=of A] {?};
        \node[state] (C) [right=of B] {$t$};
        \path [->] (A) edge (B);
        \path [->] (C) edge (B);
    \end{tikzpicture}
    \caption{Outward overlap}

    \begin {tikzpicture}[every node/.style={scale=0.8}, node distance = 1cm and 1.25cm, on grid, semithick,
        state/.style ={circle, draw, fill=black!20, text=black, minimum width = 0.75cm}]
        \node[state] (A) {$a$};
        \node[state] (B) [right=of A] {?};
        \node[state] (C) [right=of B] {$t$};
        \path [<->] (A) edge (B);
        \path [<->] (C) edge (B);
    \end{tikzpicture}
    \caption{Bidirectional overlap}
  }
  \end{subfigure}


    \caption{\textbf{Cyberbullying or not}. The leftmost thread demonstrates all five cyberbullying criteria. Although the thread in the middle contains repeated use of aggressive language, there is no harmful intent, visibility among peers, or power imbalance.
    \textbf{Overlap measures.} (right) Graphical representation of the \textbf{neighborhood overlap measures} of author $a$ and target $t$.}
    
    \label{fig:cb-and-overlap}
\end{figure*}

For instructional purposes, we provided five sample threads to demonstrate both positive and negative examples for each of the five criteria. Two of these threads are shown here. The thread in Figure~\ref{fig:threadcb} displays bullying behavior that is targeted against the green user, with all five cyberbullying criteria displayed. The thread includes repeated use of aggressive language such as ``she really fucking tried'' and ``she knows she lost.'' The bully's harmful intent is evident in the victim's defensive responses. And lastly, the thread is visible among four peers as three gang up against one, creating a power imbalance. 

The final tweet in Figure~\ref{fig:threadncb} shows the importance of context in the annotation process. If we read only this individual message, we might decide that the post is cyberbullying, but given the social context here, we can confidently assert that this post is \textit{not} cyberbullying. Although it contains the aggressive phrase ``FUCK YOU TOO BITCH'', the author does not intend harm. The message is part of a joking exchange between two friends or equals, and no other peers have joined in the conversation or interacted with the thread.

After asking workers to review these examples, we gave them a short 7-question quiz to test their knowledge. Workers were given only one quiz attempt, and they were expected to score at least 6 out of 7 questions correctly before they could proceed to the paid HIT.  Workers were then paid $\$0.12$ for each thread that they annotated. 

We successfully recruited 170 workers to label all 6,897 available threads in our dataset. They labeled an average of 121.7 threads and a median of 7 threads each. They spent an average time of 3 minutes 50 seconds, and a median time of 61 seconds per thread. For each thread, we collected annotations from three different workers, and from this data we computed our reliability metrics using Fleiss's Kappa for inter-annotator agreement as shown in Table~\ref{tab:labeled}. 

We determined ground truth for our data using a 2 out of 3 majority vote as in \citet{hosseinmardi2015analyzing}. If the message thread was missing or a \textit{target} user could not be identified, we removed the entry from the dataset, since later we would need to draw our features from both the thread and the target profile. After filtering in this way, we were left with 5,537 labeled tweets.

\begin{table}
\centering
\caption{Analysis of Labeled Twitter Data}
\resizebox{\columnwidth}{!}{%
\begin{tabular}{rrcc} 
\rowcolor{lightgray}\textbf{Criterion} & \multicolumn{1}{p{1.5cm}}{\centering \textbf{Positive} \\ \textbf{Balance}} & \multicolumn{1}{p{2.6cm}}{\centering \textbf{Inter-annotator} \\ \textbf{Agreement}} & \multicolumn{1}{p{2.5cm}}{\centering \textbf{Cyberbullying} \\ \textbf{Correlation}} \\ \hline
aggression & 74.8\% & 0.23 & 0.22 \\ \hline
repetition & 6.6\% & 0.18 & 0.27 \\ \hline
harmful intent & 16.1\% & 0.42 & 0.68 \\ \hline
visibility among peers & 30.1\% & 0.51 & 0.07 \\ \hline
target power & 34.3\% & 0.37 & 0.11 \\ \hline
author power & 3.1\% & 0.10 & -0.02 \\ \hline
equal power & 59.7\% & 0.22 & -0.09 \\ \hline
cyberbullying & 0.7\% & 0.18 & --
\end{tabular}%
\label{tab:labeled}
}
\end{table}

\subsection{Cyberbullying Transcends Cyberaggression}

As discussed earlier, some experts have argued that cyberbullying is different from online aggression~\cite{hosseinmardi2015analyzing,rosa2019automatic,salawu2017approaches}.
We asked our annotators to weigh in on this issue by asking them the subjective question for each thread: ``Based on your own intuition, is this tweet an example of cyberbullying?'' We did not use the cyberbullying label as ground truth for training models; we used this label to better understand worker perceptions of cyberbullying. We found that our workers believed cyberbullying will depend on a weighted combination of the five criteria presented in this paper, with the strongest correlate being harmful intent as shown in Table~\ref{tab:labeled}.

Furthermore, the annotators decided our dataset contained 74.8\% aggressive messages as shown in the \textit{Positive Balance} column of Table~\ref{tab:labeled}. We found that a large majority of these aggressive tweets were not labeled as ``cyberbullying.'' Rather, only 10.5\% were labeled by majority vote as cyberbullying, and only 21.5\% were considered harmful. From this data, we propose that cyberbullying and cyberaggression are not equivalent classes. Instead, cyberbullying transcends cyberaggression.



\section{Feature Engineering}
We have established that cyberbullying is a complex social phenomenon, different from the simpler notion of cyberaggression. Standard Bag of Words (BoW) features based on single sentences, such as $n$-grams and word embeddings, may thus lead machine learning algorithms to incorrectly classify friendly or joking behavior as cyberbullying \cite{hosseinmardi2015analyzing,rosa2019automatic,salawu2017approaches}. To more reliably capture the nuances of repetition, harmful intent, visibility among peers, and power imbalance, we designed a new set of features from the social and linguistic traces of Twitter users. These measures allow our classifiers to encode the dynamic relationship between the message author and target, using network and timeline similarities, expectations from language models, and other signals taken from the message thread. 

For each feature and each cyberbullying criterion, we compare the cumulative distributions of the positive and negative class using the two-sample Kolmogorov-Smirnov test. We report the Kolmogorov-Smirnov statistic $D$ (a normalized distance between the CDF of the positive and negative class) as well as the $p$-value with $\alpha = 0.05$ as our level for statistical significance.

\subsection{Text-based Features}
To construct realistic and competitive baseline models, we consider a set of standard text-based features that have been used widely throughout the literature. Specifically, we use the \texttt{NLTK} library \cite{bird2009natural} to construct unigrams, bigrams, and trigrams for each labeled message. This parallels the work of \citet{hosseinmardi2016prediction}, \citet{van2018automatic}, and \citet{xu2012learning}. Following \citet{zhang2016cyberbullying}, we incorporate counts from the Linguistic Inquiry and Word Count (LIWC) dictionary to measure the linguistic and psychological processes that are represented in the text \cite{pennebaker2007liwc2007}. We also use a modified version of the Flesch-Kincaid Grade Level and Flesch Reading Ease scores as computed in \citet{davidson2017automated}. Lastly, we encode the sentiment scores for each message using the Valence Aware Dictionary and sEntiment Reasoner (VADER) of \citet{hutto2014vader}.

\subsection{Social Network Features} \label{sec:socialnetwork}
Network features have been shown to improve text-based models \cite{huang2010analysis,singh2016cyberbullying}, and they can help classifiers distinguish between bullies and victims \cite{chelmis2017mining}. These features may also capture some of the more social aspects of cyberbullying, such as power imbalance and visibility among peers. However, many centrality measures and clustering algorithms require detailed network representations. These features may not be scalable for real-world applications. We propose a set of low-complexity measurements that can be used to encode important higher-order relations at scale. Specifically, we measure the relative positions of the author and target accounts in the directed following network by computing modified versions of Jaccard's similarity index as we now explain.

\subsubsection{Neighborhood Overlap}
Let $N^{+}(u)$ be the set of all accounts followed by user $u$ and let $N^{-}(u)$ be the set of all accounts that follow user $u$. Then $N(u) = N^{+}(u) \cup N^{-}(u)$ is the neighborhood set of $u$. We consider five related measurements of neighborhood overlap for a given author $a$ and target $t$, listed here.
\begin{eqnarray*}
   \textbf{down}(a, t) &=&  \textstyle \frac{|N^{+}(a) \cap N^{-}(t)|}{|N^{+}(a) \cup N^{-}(t)|}\\
   \textbf{up}(a, t) &=&    \textstyle \frac{|N^{-}(a) \cap N^{+}(t)|}{|N^{-}(a) \cup N^{+}(t)|}\\
   \textbf{in}(a, t) &=&    \textstyle \frac{|N^{-}(a) \cap N^{-}(t)|}{|N^{-}(a) \cup N^{-}(t)|}\\
   \textbf{out}(a, t) &=&    \textstyle \frac{|N^{+}(a) \cap N^{+}(t)|}{|N^{+}(a) \cup N^{+}(t)|}\\
   \textbf{bi}(a, t) &=&    \textstyle  \frac{|N(a) \cap N(t)|}{|N(a) \cup N(t)|}
\end{eqnarray*}
Downward overlap measures the number of two-hop paths from the author to the target along following relationships; upward overlap measures two-hop paths in the opposite direction. Inward overlap measures the similarity between the two users' follower sets, and outward overlap measures the similarity between their sets of friends. Bidirectional overlap then is a more generalized measure of social network similarity. We provide a graphical depiction for each of these features on the right side of Figure~\ref{fig:cb-and-overlap}. 

\begin{figure}
    \centering
    \begin{subfigure}{4cm}
        \centering
        \includegraphics[width=3.75cm]{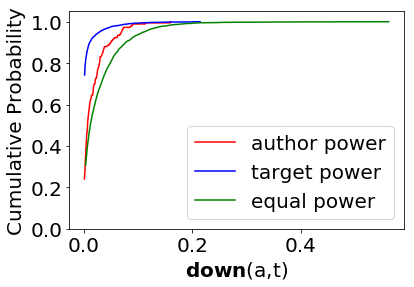}
        \caption{Downward Overlap}
    \end{subfigure}
    \begin{subfigure}{4cm}
        \centering
        \includegraphics[width=3.75cm]{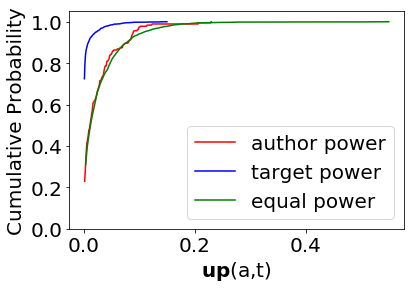}
        \caption{Upward Overlap}
    \end{subfigure}
    \begin{subfigure}{4cm}
        \centering
        \includegraphics[width=3.75cm]{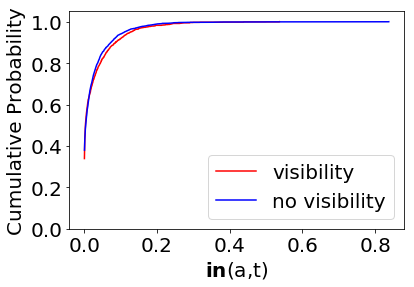}
        \caption{Inward Overlap}
    \end{subfigure}
    \begin{subfigure}{4cm}
        \centering
        \includegraphics[width=3.75cm]{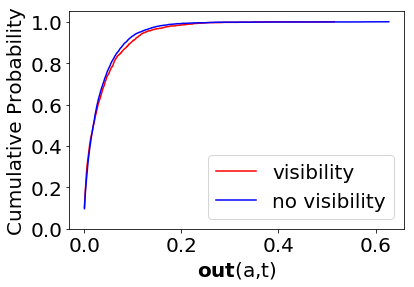}
        \caption{Outward Overlap}
    \end{subfigure}
    \caption{Cumulative Distribution Functions for \textbf{neighborhood overlap} on relevant features. These measures are shown to be predictive of \textit{power imbalance} and \textit{visibility among peers}.}
    \label{fig:cdfov}
\end{figure}

High downward overlap likely indicates that the target is socially relevant to the author, as high upward overlap indicates the author is relevant to the target. Therefore, when the author is more powerful, downward overlap is expected to be lower and upward overlap is expected be higher. This trend is slight but visible in the cumulative distribution functions of Figure~\ref{fig:cdfov} (a): downward overlap is indeed lower when the author is more powerful than when the users are equals ($D=0.143$). However, there is not a significant difference for upward overlap ($p=0.85$). We also observe that, when the target is more powerful, downward and upward overlap are both significantly lower ($D=0.516$ and $D=0.540$ respectively). It is reasonable to assume that messages can be sent to celebrities and other powerful figures without the need for common social connections. 

Next, we consider inward and outward overlap. When the inward overlap is high, the author and target could have more common visibility. Similarly, if the outward overlap is high, then the author and target both follow similar accounts, so they might have similar interests or belong to the same social circles. Both inward and outward overlaps are expected to be higher when a post is visible among peers. This is true of both distributions in Figure~\ref{fig:cdfov}. The difference in outward overlap is significant ($D=0.04$, $p=0.03$), and the difference for inward overlap is short of significant ($D=0.04$, $p=0.08$).

\subsubsection{User-based features} We also use basic user account metrics drawn from the author and target profiles. Specifically, we count the friends and followers of each user, their verified status, and the number of tweets posted within six-month snapshots of their timelines, as in  \citet{al2016cybercrime}, \citet{chatzakou2017mean}, and \citet{hosseinmardi2016prediction}.

\subsection{Timeline Features} \label{sec:timeline}
Here, we consider linguistic features, drawn from both the author and target timelines. These are intended to capture the social relationship between each user, their common interests, and the surprise of a given message relative to the author's timeline history.

\subsubsection{Message Behavior}
\begin{figure}
    \centering
    \begin{subfigure}{4cm}
        \centering
        \includegraphics[width=3.75cm]{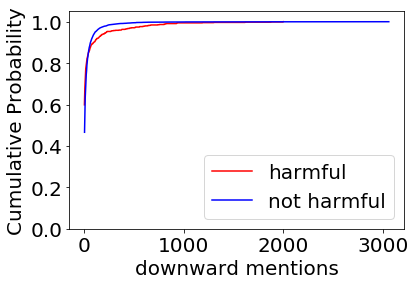}
        \caption{Downward Mentions}
    \end{subfigure}
    \begin{subfigure}{4cm}
        \centering
        \includegraphics[width=3.75cm]{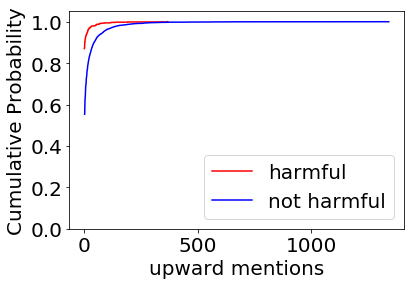}
        \caption{Upward Mentions}
    \end{subfigure}
    \begin{subfigure}{4cm}
        \centering
        \includegraphics[width=3.75cm]{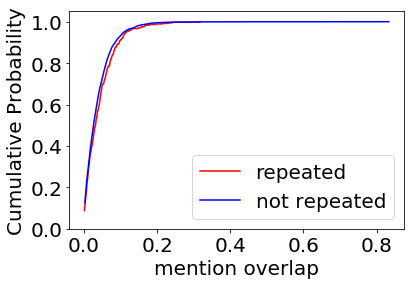}
        \caption{Mention Overlap}
    \end{subfigure}
    \begin{subfigure}{4cm}
        \centering
        \includegraphics[width=3.75cm]{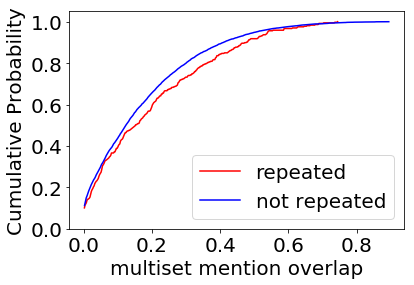}
        \caption{Multiset Mention Overlap}
    \end{subfigure}
    \caption{Cumulative Distribution Functions for \textbf{message behavior} on relevant features. These measures are shown to be indicative of \textit{harmful intent} and \textit{repetition}.}
    \label{fig:cdfmsg}
\end{figure}

To more clearly represent the social relationship between the author and target users, we consider the messages sent between them as follows:
\begin{itemize}
  \item[-] \textit{Downward mention count:} How many messages has the author sent to the target? 
  \item[-] \textit{Upward mention count:} How many messages has the target sent to the author?
  \item[-] \textit{Mention overlap:} Let $M_a$ be the set of all accounts mentioned by author $a$, and let $M_t$ be the set of all accounts mentioned by target $t$. We compute the ratio $\frac{|M_a \cap M_t|}{|M_a \cup M_t|}$.
  \item[-] \textit{Multiset mention overlap}: Let $\hat{M}_a$ be the multiset of all accounts mentioned by author $a$ (with repeats for each mention), and let $\hat{M}_t$ be the multiset of all accounts mentioned by target $t$. We measure 
  $\frac{|\hat{M}_a \cap^{*} \hat{M}_t|}{|\hat{M}_a \cup \hat{M}_t|}$ where $\cap^{*}$ takes the multiplicity of each element to be the sum of the multiplicity from $\hat{M}_a $ and the multiplicity from $\hat{M}_b$
\end{itemize}
The direct mention count measures the history of repeated communication between the author and the target. For harmful messages, downward overlap is higher ($D=0.178$) and upward overlap is lower ($D=0.374$) than for harmless messages, as shown in Figure~\ref{fig:cdfmsg}. This means malicious authors tend to address the target repeatedly while the target responds with relatively few messages. 

Mention overlap is a measure of social similarity that is based on shared conversations between the author and the target. Multiset mention overlap measures the frequency of communication within this shared space. These features may help predict visibility among peers, or repeated aggression due to pile-on bullying situations. We see in Figure~\ref{fig:cdfmsg} that repeated aggression is linked to slightly greater mention overlap ($D=0.07$, $p=0.07$), but the trend is significant only for multiset mention overlap ($D=0.08$, $p=0.03$).

\subsubsection{Timeline Similarity}
Timeline similarity is used to indicate common interests and shared topics of conversation between the author and target timelines. High similarity scores might reflect users' familiarity with one another, or suggest that they occupy similar social positions. This can be used to distinguish cyberbullying from harmless banter between friends and associates. To compute this metric, we represent the author and target timelines as TF-IDF vectors $\vec{A}$ and $\vec{T}$. We then take the \textbf{cosine similarity} between the vectors as
\begin{equation*}
    \cos{\theta} = \frac{\vec{A} \cdot \vec{T}}{\|\vec{A}\|\|\vec{T}\|}.
\end{equation*}
A cosine similarity of 1 means that users' timelines had identical counts across all weighted terms; a cosine similarity of 0 means that their timelines did not contain any words in common. We expect higher similarity scores between friends and associates. 

\begin{figure}
    \centering
    \begin{subfigure}{4cm}
        \centering
        \includegraphics[width=3.75cm]{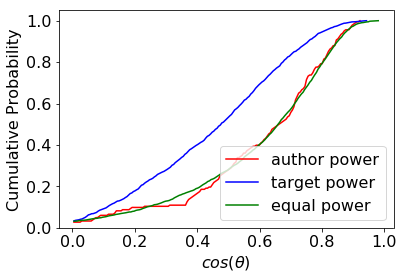}
        \caption{Timeline Similarity}
    \end{subfigure}
    \begin{subfigure}{4cm}
        \centering
        \includegraphics[width=3.75cm]{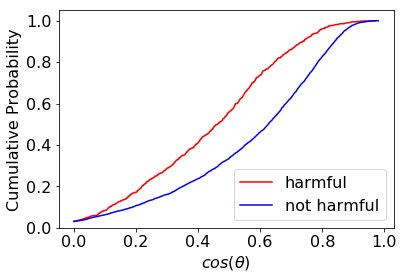}
        \caption{Timeline Similarity}
    \end{subfigure}
    \caption{Cumulative Distribution Functions for \textbf{timeline similarity} on relevant features. These measures are shown to be predictive of \textit{power imbalance} and \textit{harmful intent}.}
    \label{fig:cdfsim}
\end{figure}

In Figure~\ref{fig:cdfsim} (a), we see that the timelines were significantly less similar when the target was in a position of greater power ($D=0.294$). This is not surprising, since power can be derived from such differences between social groups. We do not observe the same dissimilarity when the author was more powerful ($p=0.58$). What we do observe is likely caused by noise from extreme class imbalance and low inter-annotator agreement on labels for author power. 

Turning to Figure~\ref{fig:cdfsim} (b), we see that aggressive messages were less likely to harbor harmful intent if they were sent between users with similar timelines ($D=0.285$). Aggressive banter between friends is generally harmless, so again, this confirms our intuitions.

\begin{figure}
    \centering
    \begin{subfigure}{4cm}
        \centering
        \includegraphics[width=3.75cm]{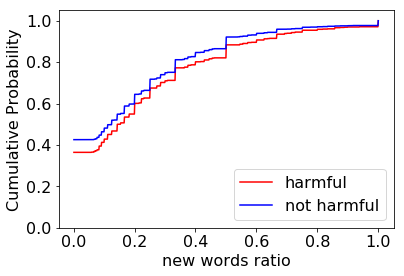}
        \caption{New Words Ratio}
    \end{subfigure}
    \begin{subfigure}{4cm}
        \centering
        \includegraphics[width=3.75cm]{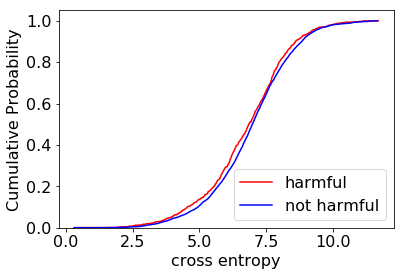}
        \caption{Cross Entropy}
    \end{subfigure}
    \caption{Cumulative Distribution Functions for \textbf{language models} on relevant features. These measures are shown to be predictive of \textit{harmful intent}.}
    \label{fig:cdflang}
\end{figure}

\subsubsection{Language Models}
Harmful intent is difficult to measure in isolated messages because social context determines pragmatic meaning. We attempt to approximate the author's harmful intent by measuring the linguistic ``surprise'' of a given message relative to the author's timeline history. We do this in two ways: through a simple ratio of new words, and through the use of language models. 

To estimate historical language behavior, we count unigram and bigram frequencies from a 4-year snapshot of the author's timeline. Then, after removing all URLs, punctuation, stop words, mentions, and hashtags from the original post, we take the cardinality of the set unigrams in the post having zero occurrences in the timeline. Lastly, we divide this count by the length of the processed message to arrive at our \textbf{new words ratio}. We can also build a language model from the bigram frequencies, using Kneser-Ney smoothing as implemented in {\fontfamily{qcr}\selectfont NLTK} \cite{bird2009natural}. From the language model, we compute the surprise of the original message $m$ according to its \textbf{cross-entropy}, given by
\begin{equation*}
    \text{H}(m) = -\frac{1}{N} \sum_{i=1}^N \log{P(b_i)}
\end{equation*}
where $m$ is composed of bigrams $b_1, b_2, \dots, b_N$, and $P(b_i)$ is the probability of the $i$th bigram from the language model. 

We see in Figure~\ref{fig:cdflang} that harmfully intended messages have a greater density of new words ($D=0.06$). This is intuitive, since attacks may be staged around new topics of conversation. However, the cross entropy of these harmful messages is slightly lower than for harmless messages ($D=0.06$). This may be due to harmless jokes, since joking messages might depart more from the standard syntax of the author's timeline.

\subsection{Thread Features}
Finally, we turn to the messages of the thread itself to compute measures of visibility and repeated aggression. 

\subsubsection{Visibility}
To determine the public visibility of the author's post, we collect basic measurements from the interactions of other users in the thread. They are as follows.
\begin{itemize}
    \item[-] \textit{Message count}: Count the messages posted in the thread
    \item[-] \textit{Reply message count}: Count the replies posted in the thread after the author's first comment.
    \item[-] \textit{Reply user count}: Count the users who posted a reply in the thread after the author's first comment.
    \item[-] \textit{Maximum author favorites}: The largest number of favorites the author received on a message in the thread.
    \item[-] \textit{Maximum author retweets}: The largest number of retweets the author received on a message in the thread.
\end{itemize}

\begin{table}
\centering
\caption{Feature Combinations}
\resizebox{\columnwidth}{!}{%
\begin{tabular}{rccccc} \hline
\rowcolor{lightgray}\textit{Feature} & \textit{BoW} & \textit{Text} & \textit{User} & \textit{Proposed} &  \textit{Combined} \\ \hline
$n$-grams & \ding{51} & \ding{51} & & & \ding{51} \\ \hline
LIWC, VADER, Flesch-Kincaid & & \ding{51} & & & \ding{51} \\ \hline
Friend/following counts, tweet count, verified & & & \ding{51} & \ding{51} & \ding{51} \\ \hline
Neighborhood overlap measures & & & \ding{51} & \ding{51} & \ding{51} \\ \hline
Mention counts and overlaps & & & \ding{51} & \ding{51} & \ding{51} \\ \hline
Timeline similarity & & & \ding{51} & \ding{51} & \ding{51} \\ \hline
New words ratio, cross-entropy & & & \ding{51} & \ding{51} & \ding{51} \\ \hline
Thread visibility features & & & &  \ding{51} & \ding{51} \\ \hline
Thread aggression features & & & & \ding{51} & \ding{51} \\ \hline
\end{tabular}%
\label{tab:featurecombos}
}
\end{table}

\begin{table}
\centering
\caption{Precision}
\resizebox{\columnwidth}{!}{%
\begin{tabular}{rccccc} \hline
\rowcolor{lightgray}\textit{Criterion} & \textit{BoW} & \textit{Text} & \textit{User} & \textit{Proposed} &  \textit{Combined} \\ \hline
aggression & 82.5\% & 82.3\% & 77.1\% & 78.7\% & \textbf{82.6\%} \\ \hline
repetition & 7.8\% & 13.4\% & 7.7\% & 15.3\% & \textbf{31.7\%}\\ \hline
harmful intent & 29.6\% & 49.4\% & 35.8\% & 34.5\% & \textbf{55.3\%} \\ \hline
visibility among peers & 30.8\% & 34.3\% & 34.0\% & 42.2\% & \textbf{46.8\%} \\ \hline
author power & 1.9\% & 3.6\% & 7.6\% & 9.8\% & \textbf{17.0\%} \\ \hline
target power & 43.5\% & 51.5\% & \textbf{77.6\%} & 75.2\% & 77.0\% \\ \hline
\end{tabular}%
\label{tab:precision}
}
\end{table}

\subsubsection{Aggression}
To detect repeated aggression, we again employ the hate speech and offensive language classifier of \citet{davidson2017automated}. Each message is given a binary label according to the classifier-assigned class: aggressive (classified as hate speech or offensive language) or non-aggressive (classified as neither hate speech nor offensive language). From these labels, we derive the following features.
\begin{itemize}
    \item[-] \textit{Aggressive message count}: Count the messages in the thread classified as aggressive
    \item[-] \textit{Aggressive author message count}: Count the author's messages that were classified as aggressive
    \item[-] \textit{Aggressive user count}: Of the users who posted a reply in the thread after the author first commented, count how many had a message classified as aggressive
\end{itemize}

\section{Experimental Evaluation}

Using our proposed features from the previous section and ground truth labels from our annotation task, we trained a separate Logistic Regression classifier for each of the five cyberbullying criteria, and we report precision, recall, and $F_1$ measures over each binary label independently. We averaged results using five-fold cross-validation, with 80\% of the data allocated for training and 20\% of the data allocated for testing at each iteration. To account for the class imbalance in the \emph{training} data, we used the synthetic minority over-sampling technique (SMOTE) \cite{chawla2002smote}. We did not over-sample testing sets, however, to ensure that our tests better match the class distributions obtained as we did by pre-filtering for aggressive directed Twitter messages.

We compare our results across the five different feature combinations given in Table~\ref{tab:featurecombos}. Note that because we do not include thread features in the \textit{User} set, it can be used for cyberbullying prediction and early intervention. The \textit{Proposed} set can be used for detection, sinct it is a collection of all newly proposed features, including thread features. The \textit{Combined} adds these to the baseline text features.

\begin{table}
\centering
\caption{Recall}
\resizebox{\columnwidth}{!}{%
\begin{tabular}{rccccc} \hline
\rowcolor{lightgray}\textit{Criterion} & \textit{BoW} & \textit{Text} & \textit{User} & \textit{Proposed} &  \textit{Combined} \\ \hline
aggression & 77.0\% & 84.8\% & 47.8\% & 51.6\% & \textbf{85.6\%} \\ \hline
repetition & 17.6\% & 7.3\% & 49.5\% & \textbf{64.3\%} & 26.2\% \\ \hline
harmful intent & 40.2\% & 44.4\% & 63.4\% & \textbf{67.7\%} & 52.7\% \\ \hline
visibility among peers & 34.8\% & 20.4\% & 47.1\% & \textbf{54.2\%} & 33.7\% \\ \hline
author power & 6.5\% & 1.6\% & 74.1\% & \textbf{80.0\%} & 11.9\% \\ \hline
target power & 49.4\% & 43.3\% & 73.3\% & \textbf{80.8\%} & 71.1\% \\ \hline
\end{tabular}%
\label{tab:recall}
}
\end{table}

\begin{table}
\centering
\caption{$F_1$ Scores}
\resizebox{\columnwidth}{!}{%
\begin{tabular}{rccccc} \hline
\rowcolor{lightgray}\textit{Criterion} & \textit{BoW} & \textit{Text} & \textit{User} & \textit{Proposed} &  \textit{Combined} \\ \hline
aggression & 79.7\% & 83.5\% & 59.0\% & 62.3\% & \textbf{84.1\%} \\ \hline
repetition & 10.8\% & 9.4\% & 13.3\% & 24.7\% & \textbf{28.7\%} \\ \hline
harmful intent & 34.1\% & 46.7\% & 38.7\% & 45.7\% & \textbf{53.8\%} \\ \hline
visibility among peers & 32.7\% & 25.5\% & 39.5\% & \textbf{47.4\%} & 45.5\% \\ \hline
author power & 2.9\% & 2.2\% & 13.7\% & \textbf{17.5\%} & 14.0\% \\ \hline
target power & 46.2\% & 47.0\% & 75.3\% & \textbf{77.9\%} & 73.9\% \\ \hline
\end{tabular}%
\label{tab:f1}
}
\end{table}

The performance of the different classifiers is summarized in Tables~\ref{tab:precision}, \ref{tab:recall}, and \ref{tab:f1}.
Here, we see that Bag of Words and text-based methods performed well on the aggressive language classification task, with an $F_1$ score of 83.5\%. This was expected and the score aligns well with the success of other published results of Table~\ref{tab:sota}. 

Cyberbullying detection is more complex than simply identifying aggressive text, however. We find that these same baseline methods fail to reliably detect repetition, harmful intent, visibility among peers, and power imbalance, as shown by the low recall scores in Table~\ref{tab:recall}. We conclude that our investigation of socially informed features was justified.

Our proposed set of features beats recall scores for lexically trained baselines in all but the aggression criterion. We also improve precision scores for repetition, visibility among peers, and power imbalance. When we combine all features, we see our $F_1$ scores beat baselines for each criterion. This demonstrates the effectiveness of our approach, using linguistic similarity and community measurements to encode social characteristics for cyberbullying classification. 

Similar results were obtained by replacing our logistic regression model with any of a random forest model, support vector machine (SVM), AdaBoost, or Multilayer Perceptron (MLP). We report all precision, recall, and $F_1$ scores in Appendix 2, Tables~\ref{tab:rf_precision}-\ref{tab:mlp_f1}. We chose to highlight logistic regression because it can be more easily interpreted. As a result, we can identify the relative importance of our proposed features. The feature weights are also given in Appendix 2, Tables~\ref{tab:weights_aggr}-\ref{tab:weights_power}. There we observe a trend. The \textit{aggressive language} and \textit{repetition} criteria are dominated by lexical features; the \textit{harmful intent} is split between lexical and historical communication features; and the \textit{visibility among peers} and \textit{target power} criteria are dominated by our proposed social features. 

Although we achieve moderately competitive scores in most categories, our classifiers are still over-classifying cyberbullying cases. Precision scores are generally much lower than recall scores across all models. To reduce our misclassification of false positives and better distinguish between joking or friendly banter and cyberbullying, it may be necessary to mine for additional social features. Overall, we should work to increase all $F_1$ scores to above 0.8 before we can consider our classifiers ready for real-world applications \cite{rosa2019automatic}. 

\section{Discussion}
\subsubsection{Limitations}
Our study focuses on the Twitter ecosystem and a small part of its network. 
The initial sampling of tweets was based on a machine learning classifier of aggressive English language. This classifier has an F1 score of 0.90~\cite{davidson2017automated}. Even with this filter, only 0.7\% of tweets were deemed by a majority of MTurk workers as cyberbullying (Table~\ref{tab:labeled}). This extreme class imbalance can disadvantage a wide range of machine learning models.
Moreover, the MTurk workers exhibited only moderate inter-annotator agreement (Table~\ref{tab:labeled}). 
We also acknowledge that notions of \textit{harmful intent} and \textit{power imbalance} can be subjective, since they may depend on the particular conventions or social structure of a given community. For these reasons, we recognize that cyberbullying still has not been unambiguously defined. Moreover, their underlying constructs are difficult to identify. In this study, we did not train workers to recognize subtle cues for interpersonal popularity, nor the role of anonymity in creating a power imbalance. 

Furthermore, because we lack the authority to define cyberbullying, we cannot assert a two-way implication between cyberbullying and the five criteria outlined here. It may be possible for cyberbullying to exist with only one criterion present, such as harmful intent. 
Our five criteria also might not span all of the dimensions of cyberbullying. However, they are representative of the literature in both the social science and machine learning communities, and they can be used in weighted combinations to accommodate new definitions. 

The main contribution of our paper is not that we solved the problem of cyberbullying detection. Instead, we have exposed the challenge of defining and measuring cyberbullying activity, which has been historically overlooked in the research community.

\subsubsection{Future Directions}
Cyberbullying detection is an increasingly important and yet challenging problem to tackle. A lack of detailed and appropriate real-world datasets stymies progress towards more reliable detection methods. With cyberbullying being a systemic issue across social media platforms, we urge the development of a methodology for data sharing with researchers that provides adequate access to rich data to improve on the early detection of cyberbullying while also addressing the sensitive privacy issues that accompany such instances.



\section{Conclusion}
In this study, we produced an original dataset for cyberbullying detection research and an approach that leverages this dataset to more accurately detect cyberbullying. Our labeling scheme was designed to accommodate the cyberbullying definitions that have been proposed throughout the literature. In order to more accurately represent the nature of cyberbullying, we decomposed this complex issue into five representative characteristics. Our classes distinguish cyberbullying from other related behaviors, such as isolated aggression or crude joking. To help annotators infer these distinctions, we provided them with the full context of each message's reply thread, along with a list of the author's most recent mentions. In this way, we secured a new set of labels for more reliable cyberbullying representations.

From these ground truth labels, we designed a new set of features to quantify each of the five cyberbullying criteria. Unlike previous text-based or user-based features, our features measure the relationship between a message author and target. We show that these features improve the performance of standard text-based models. These results demonstrate the relevance of social-network and language-based measurements to account for the nuanced social characteristics of cyberbullying.

Despite improvements over baseline methods, our classifiers have not attained the high levels of precision and recall that should be expected of real-world detection systems. For this reason, we argue that the challenging task of cyberbullying detection remains an open research problem.

\section*{Acknowledgements}
This material is based upon work supported by the Defense Advanced Research Projects Agency (DARPA) under Agreement No. HR0011890019, and by the National Science Foundation (NSF) under Grant No. 1659886 and Grant No. 1553579. 

\bibliography{refs}
\bibliographystyle{aaai}

\section{Appendix 1: Analysis of the Real-World Class Distribution for Cyberbullying Criteria}

To understand the real-world class distribution for the cyberbullying criteria, we randomly selected 222 directed English tweets from an unbiased sample of drawn from the Twitter Decahose stream across the entire month of October 2016. Using the same methodology given in the paper, we had these tweets labeled three times each on Amazon Mechanical Turk. Again, ground truth was determined using 2 out of 3 majority vote. Upon analysis, we found that the positive class balance was prohibitively small, especially for \textit{repetition}, \textit{harmful intent}, \textit{visibility among peers}, and \textit{author power}, which were all under 5\%. 


\begin{table}[h!]
\centering
\caption{Analysis of Unfiltered Decahose Data}
\resizebox{\columnwidth}{!}{%
\begin{tabular}{rccc} \hline
\rowcolor{lightgray}\textbf{Criterion} & \multicolumn{1}{|p{1.5cm}|}{\centering \textbf{Positive} \\ \textbf{Balance}} & \multicolumn{1}{|p{2.6cm}|}{\centering \textbf{Inter-annotator} \\ \textbf{Agreement}} & \multicolumn{1}{|p{2.5cm}|}{\centering \textbf{Cyberbullying} \\ \textbf{Correlation}} \\ \hline
aggression & 6.3\% & 0.23 & 0.68 \\ \hline
repetition & 0.9\% & 0.04 & 0.46 \\ \hline
harmful intent & 1.4\% & 0.31 & 0.75 \\ \hline
visibility among peers & 0.17\% & 0.51 & 0.11 \\ \hline
target power & 22.5\% & 0.23 & 0.11 \\ \hline
author power & 3.6\% & 0.04 & 0.06 \\ \hline
equal power & 64.7\% & 0.15 & -0.14 \\ \hline
cyberbullying & 2.7\% & 0.25 & - \\ \hline
\end{tabular}%
\label{tab:decahose_data}
}
\end{table}

\section{Appendix 2: Model Evaluation}

For the sake of comparison, we provide precision, recall, and $F_1$ scores for five different machine learning models: $k$-nearest neighbors (KNN), random forest, support vector machine (SVM), AdaBoost, and Multilayer Perceptron (MLP). Then we provide feature weights for our logistic regression model trained on each of the five cyberbullying criteria.



\begin{table}[h!]
\centering
\caption{Random Forest Precision}
\resizebox{\columnwidth}{!}{%
\begin{tabular}{rrrrrrr} \hline
\rowcolor{lightgray}\textit{Criterion} & \textit{BoW} & \textit{Text} & \textit{User} & \textit{Proposed} &  \textit{Combined} \\ \hline
aggression             &  77.6\% &  \textbf{80.1\%} &  78.3\% &    78.7\% &    79.7\% \\ \hline
repetition             &   6.5\% &   6.8\% &   7.7\% &    \textbf{16.1\%} &    10.8\% \\ \hline
harmful intent         &  18.4\% &  28.1\% &  33.2\% &    33.4\% &    \textbf{43.1\%} \\ \hline
visibility among peers &  28.7\% &  32.7\% &  34.8\% &    \textbf{42.8\%} &    35.1\% \\ \hline
target power           &  39.3\% &  43.3\% &  \textbf{77.9\%} &    74.5\% &    69.6\% \\ \hline
\end{tabular}%
\label{tab:rf_precision}
}
\end{table}


\begin{table}[h!]
\centering
\caption{AdaBoost Precision}
\resizebox{\columnwidth}{!}{%
\begin{tabular}{rrrrrrr} \hline
\rowcolor{lightgray}\textit{Criterion} & \textit{BoW} & \textit{Text} & \textit{User} & \textit{Proposed} &  \textit{Combined} \\ \hline
aggression             &  \textbf{82.6\%} &  81.6\% &  77.0\% &    77.5\% &    81.6\% \\ \hline
repetition             &   7.8\% &   9.0\% &   7.3\% &    16.6\% &    \textbf{25.8\%}\\ \hline
harmful intent         &  29.1\% &  46.4\% &  34.3\% &    39.9\% &    \textbf{60.0\%} \\ \hline
visibility among peers &  30.5\% &  32.9\% &  35.9\% &    45.8\% &    \textbf{46.1\%} \\ \hline
target power           &  42.5\% &  46.5\% &  78.0\% &    \textbf{78.2\%} &    77.9\% \\ \hline
\end{tabular}%
\label{tab:ada_precision}
}
\end{table}

\begin{table} [h!]
\centering
\caption{MLP Precision}
\resizebox{\columnwidth}{!}{%
\begin{tabular}{rrrrrrr} \hline
\rowcolor{lightgray}\textit{Criterion} & \textit{BoW} & \textit{Text} & \textit{User} & \textit{Proposed} &  \textit{Combined} \\ \hline
aggression             &  \textbf{82.8\%} &  78.8\% &  76.7\% &    77.4\% &    78.3\% \\ \hline
repetition             &   7.7\% &   8.7\% &   8.6\% &    16.9\% &    \textbf{19.6\%} \\ \hline
harmful intent         &  27.4\% &  42.8\% &  37.3\% &    38.4\% &    \textbf{46.8\%} \\ \hline
visibility among peers &  30.1\% &  34.0\% &  34.3\% &    \textbf{41.6\%} &    38.5\% \\ \hline
target power           &  39.6\% &  45.2\% &  \textbf{74.3\%} &    72.0\% &    68.6\% \\ \hline
\end{tabular}%
\label{tab:mlp_precision}
}
\end{table}



\begin{table}[h!]
\centering
\caption{Random Forest Recall}
\resizebox{\columnwidth}{!}{%
\begin{tabular}{rrrrrrr} \hline
\rowcolor{lightgray}\textit{Criterion} & \textit{BoW} & \textit{Text} & \textit{User} & \textit{Proposed} &  \textit{Combined} \\ \hline
aggression             &  56.4\% &  \textbf{78.5\%} &  43.7\% &    45.3\% &    76.2\% \\ \hline
repetition             &  36.2\% &  24.9\% &  46.3\% &    \textbf{64.7\%} &    29.9\% \\ \hline
harmful intent         &  42.4\% &  35.1\% &  \textbf{78.4\%} &    78.2\% &    53.5\% \\ \hline
visibility among peers &  48.1\% &  30.6\% &  \textbf{50.5\%} &    49.9\% &    32.5\% \\ \hline
target power           &  60.1\% &  38.0\% &  79.0\% &    \textbf{81.9\%} &    76.7\% \\ \hline
\end{tabular}%
\label{tab:rf_recall}
}
\end{table}


\begin{table}[h!]
\centering
\caption{AdaBoost Recall}
\resizebox{\columnwidth}{!}{%
\begin{tabular}{rrrrrrr} \hline
\rowcolor{lightgray}\textit{Criterion} & \textit{BoW} & \textit{Text} & \textit{User} & \textit{Proposed} &  \textit{Combined} \\ \hline
aggression             &  75.0\% &  \textbf{86.4\%} &  65.9\% &    77.4\% &    86.3\% \\ \hline
repetition             &  23.8\% &   4.1\% &  26.8\% &    \textbf{31.2\%} &    17.8\% \\ \hline
harmful intent         &  44.4\% &  37.8\% &  \textbf{57.0\%} &    52.8\% &    50.8\% \\ \hline
visibility among peers &  41.0\% &  15.4\% &  42.8\% &    \textbf{43.1\%} &    32.0\% \\ \hline
target power           &  56.0\% &  39.4\% &  \textbf{81.8\%} &    81.0\% &    75.6\% \\ \hline
\end{tabular}%
\label{tab:ada_recall}
}
\end{table}

\begin{table}[h!]
\centering
\caption{MLP Recall}
\resizebox{\columnwidth}{!}{%
\begin{tabular}{rrrrrrr} \hline
\rowcolor{lightgray}\textit{Criterion} & \textit{BoW} & \textit{Text} & \textit{User} & \textit{Proposed} &  \textit{Combined} \\ \hline
aggression             &  64.1\% &  \textbf{86.5\%} &  65.5\% &    68.0\% &    85.6\% \\ \hline
repetition             &  26.8\% &   6.8\% &  22.5\% &   \textbf{27.1\%} &    12.6\% \\ \hline
harmful intent         &  51.0\% &  33.3\% &  \textbf{57.0\%} &    \textbf{57.0\%} &    37.2\% \\ \hline
visibility among peers &  51.6\% &  23.5\% &  45.6\% &    \textbf{50.2\%} &    26.5\% \\ \hline
target power           &  61.6\% &  37.5\% &  \textbf{76.5\%} &    76.2\% &    65.6\% \\ \hline
\end{tabular}%
\label{tab:mlp_recall}
}
\end{table}



\begin{table}[h!]
\centering
\caption{Random Forest $F_1$}
\resizebox{\columnwidth}{!}{%
\begin{tabular}{rrrrrrr} \hline
\rowcolor{lightgray}\textit{Criterion} & \textit{BoW} & \textit{Text} & \textit{User} & \textit{Proposed} &  \textit{Combined} \\ \hline
aggression             &  65.2\% &  \textbf{79.3\%} &  56.0\% &    57.5\% &    77.9\% \\ \hline
repetition             &  11.0\% &  10.6\% &  13.2\% &    \textbf{25.8\%} &    15.8\% \\ \hline
harmful intent         &  25.6\% &  31.1\% &  46.6\% &    46.8\% &    \textbf{47.7\%} \\ \hline
visibility among peers &  35.7\% &  30.8\% &  41.2\% &    \textbf{46.1\%} &    33.6\% \\ \hline
target power           &  47.4\% &  39.9\% &  \textbf{78.4\%} &    78.0\% &    72.8\% \\ \hline
\end{tabular}%
\label{tab:rf_f1}
}
\end{table}


\begin{table}[h!]
\centering
\caption{AdaBoost $F_1$}
\resizebox{\columnwidth}{!}{%
\begin{tabular}{rrrrrrr} \hline
\rowcolor{lightgray}\textit{Criterion} & \textit{BoW} & \textit{Text} & \textit{User} & \textit{Proposed} &  \textit{Combined} \\ \hline
aggression             &  78.6\% &  \textbf{83.9\%} &  71.0\% &    77.5\% &    \textbf{83.9\%} \\ \hline
repetition             &  11.7\% &   5.6\% &  11.5\% &    \textbf{21.6\%} &    20.9\% \\ \hline
harmful intent         &  35.1\% &  41.6\% &  42.8\% &    45.4\% &    \textbf{55.0\%} \\ \hline
visibility among peers &  34.9\% &  21.0\% &  39.1\% &    \textbf{44.3\%} &    37.8\% \\ \hline
target power           &  48.3\% &  42.7\% &  \textbf{79.8\%} &    79.6\% &    76.7\% \\ \hline
\end{tabular}%
\label{tab:ada_f1}
}
\end{table}

\begin{table}[th!]
\centering
\caption{MLP $F_1$}
\resizebox{\columnwidth}{!}{%
\begin{tabular}{rrrrrrr} \hline
\rowcolor{lightgray}\textit{Criterion} & \textit{BoW} & \textit{Text} & \textit{User} & \textit{Proposed} &  \textit{Combined} \\ \hline
aggression             &  72.2\% &  \textbf{82.5\%} &  70.7\% &    72.4\% &    81.8\% \\ \hline
repetition             &  12.0\% &   7.6\% &  12.4\% &    \textbf{20.7\%} &    15.2\% \\ \hline
harmful intent         &  35.7\% &  37.3\% &  45.0\% &    \textbf{45.8\%} &    41.3\% \\ \hline
visibility among peers &  38.0\% &  27.7\% &  39.2\% &    \textbf{45.5\%} &    31.4\% \\ \hline
target power           &  48.2\% &  41.0\% &  \textbf{75.4\%} &    74.0\% &    67.0\% \\ \hline
\end{tabular}%
\label{tab:mlp_f1}
}
\end{table}

\begin{table}[h!]
\centering
\caption{Top Absolute Weights for Aggressive Language}
\resizebox{0.6\columnwidth}{!}{%
    \begin{tabular}{rrr} \hline
    \rowcolor{lightgray}\textit{Rank} & \textit{Feature} & \textit{Weight} \\ \hline
    1 & affect (LIWC) & -1.34 \\
    2 & sexual (LIWC) & 1.07 \\
    3 & negemo (LIWC) & 0.90 \\
    4 & maximum author retweets & 0.86 \\
    5 & relativ (LIWC) & -0.75 \\
    6 & bio (LIWC) & -0.69 \\
    7 & posemo (LIWC) & 0.66 \\
    8 & num chars & -0.64 \\
    9 & space (LIWC) & 0.52 \\
    10 & upward overlap & 0.51
    \end{tabular}
    \label{tab:weights_aggr}
}

\bigskip

\centering
\caption{Top Absolute Weights for Repetition Features}
\resizebox{0.6\columnwidth}{!}{%
    \begin{tabular}{rrr} \hline
    \rowcolor{lightgray}\textit{Rank} & \textit{Feature} & \textit{Weight} \\ \hline
    1 & negemo (LIWC) & 1.40 \\
    2 & author verified status & -1.32 \\
    3 & affect (LIWC) & -1.24 \\
    4 & cogmech (LIWC) & -0.96 \\
    5 & relativ (LIWC) & -0.89 \\
    6 & posemo (LIWC) & 0.80 \\ 
    7 & social (LIWC) & 0.77 \\
    8 & aggressive user count & 0.63 \\
    9 & upward overlap & 0.62 \\
    10 & number of unique terms & 0.61 \\
    \end{tabular}
    \label{tab:weights_rep}
}

\bigskip

\centering
\caption{Top Absolute Weights for Harmful Intent}
\resizebox{0.6\columnwidth}{!}{%
    \begin{tabular}{rrr} \hline
    \rowcolor{lightgray}\textit{Rank} & \textit{Feature} & \textit{Weight} \\ \hline
    1 & number of words & -1.70 \\
    2 & number of unique terms & 1.41 \\
    3 & bio (LIWC) & -1.05 \\
    4 & funct (LIWC) & 0.95 \\
    5 & author follower count & -0.90 \\
    6 & present (LIWC) & 0.83 \\
    7 & you (LIWC) & 0.83 \\
    8 & message count & 0.79 \\
    9 & upward mention count & -0.71 \\
    10 & verb (LIWC) & -0.67 \\
    \end{tabular}
    \label{tab:weights_harm}
}

\bigskip

\centering
\caption{Top Absolute Weights for Visibility Among Peers}
\resizebox{0.6\columnwidth}{!}{%
    \begin{tabular}{rrr} \hline
    \rowcolor{lightgray}\textit{Rank} & \textit{Feature} & \textit{Weight} \\ \hline
    1 & author follower count & 6.29 \\
    2 & maximum author retweets & -1.63 \\
    3 & maximum author favorites & 1.46 \\
    4 & aggressive user count & -1.36 \\
    5 & number of words & -1.16 \\
    6 & reply user count & 1.03 \\
    7 & number of unique terms & 1.02 \\
    8 & reply message count & -0.91 \\
    9 & message count & 0.77 \\
    10 & affect (LIWC) & -0.67
    \end{tabular}
    \label{tab:weights_peer}
}

\bigskip

\centering
\caption{Top Absolute Weights for Target Power}
\resizebox{0.6\columnwidth}{!}{%
    \begin{tabular}{rrr} \hline
    \rowcolor{lightgray}\textit{Rank} & \textit{Feature} & \textit{Weight} \\ \hline
    1 & target follower count & 2.28 \\
    2 & author follower count & -1.67 \\
    3 & bidirectional overlap & -1.22 \\
    4 & target verified status & 1.20 \\
    5 & upward overlap & -1.11 \\
    6 & downward overlap & 1.04 \\
    7 & relativ (LIWC) & 0.76 \\
    8 & reply user count & -0.69 \\
    9 & space (LIWC) & -0.68 \\
    10 & message count & -0.63 \\
    \end{tabular}
    \label{tab:weights_power}
}
\end{table}

\end{document}